\documentclass[conference]{IEEEtran}

\usepackage[cmex10]{amsmath}
\usepackage{amssymb}
\usepackage{amsfonts}
\usepackage{amsthm}
\usepackage{graphicx}
\usepackage{mathrsfs}
\usepackage{cite}
\usepackage[all]{xy}
\usepackage{multirow}
\usepackage{mathdots}

\newcommand{\leqdot}{\mathrel{\vcenter{\offinterlineskip
\hbox{\hskip1ex $\cdot$}\vskip-1.27ex\hbox{$\leqslant$}}}}

\DeclareMathOperator{\rank}{rank}
\DeclareMathOperator*{\argmin}{arg\,min}
\DeclareMathOperator{\trans}{T}
\newcommand{\defend}{$\hfill\IEEEQEDclosed$}

\theoremstyle{definition}
\newtheorem{thm}{Theorem}[]
\newtheorem{cor}[thm]{Corollary}
\newtheorem{lemma}[thm]{Lemma}
\newtheorem{prop}[thm]{Proposition}

\theoremstyle{definition}
\newtheorem*{defn}{Definition}

\begin{document}
\title{Fountain Codes with Varying Probability Distributions}
\author{\IEEEauthorblockN{Kai~Fong~Ernest~Chong, Ernest~Kurniawan, Sumei~Sun and Kai~Yen}
\IEEEauthorblockA{Institute for Infocomm Research, A*Star\\ 1 Fusionopolis Way, \#21-01 Connexis (South Tower), Singapore 138632\\ Email: \{kfchong,ekurniawan,sunsm,yenkai\}@i2r.a-star.edu.sg}}

\maketitle
\begin{abstract}
Fountain codes are rateless erasure-correcting codes, i.e., an essentially infinite stream of encoded packets can be generated from a finite set of data packets. Several fountain codes have been proposed recently to minimize overhead, many of which involve modifications of the Luby transform (LT) code. These fountain codes, like the LT code, have the implicit assumption that the probability distribution is fixed throughout the encoding process. In this paper, we will use the theory of posets to show that this assumption is unnecessary, and by dropping it, we can achieve overhead reduction by as much as 64\% lower than LT codes. We also present the fundamental theory of probability distribution designs for fountain codes with non-constant probability distributions that minimize overhead.
\end{abstract}

\section{Introduction}
Since the introduction of the Luby transform (LT) code \cite{LTcodes}, the first practical realization of a fountain code\cite{fountain}, other fountain code designs have surfaced. Examples include the raptor code\cite{Raptor}, which involves precoding an LT code with an outer erasure-correcting code, the real-time oblivious code\cite{RT-jnl}, which has a low memory requirement but which extensively uses a feedback channel and has a larger overhead than the LT code \cite[Fig. 3]{shifting2}, the systematic LT code\cite{systematic}, which yields low overhead using soft decoding, and the reconfigurable rateless code\cite{reconfigurable-jnl}, which varies block length and encoding strategy, but which relies on frequent feedback so as to achieve overhead reduction. These new designs employ modifications of LT codes or rely on a feedback channel to minimize overhead, so in the absence of a feedback channel, existing fountain codes have the implicit assumption that the probability distribution is fixed throughout the generation of the output symbols.

In this paper, we will show that by dropping this assumption, we can achieve significant overhead reduction while maintaining the same encoding and decoding complexities. In contrast to current fountain codes having fixed probability distributions, we consider fountain codes with non-constant probability distributions, without the assumption of any feedback channel, although feedback for acknowledgement is still required.

Arguments in this paper assume a familiarity with both probability theory (see, e.g., \cite{probability}) and the theory of posets (see, e.g., \cite[Ch. 3]{stanley}). For any poset $\mathbb{P}$ and any $a,b\in \mathbb{P}$, denote $a \lessdot b$ to mean $b$ covers $a$, and denote $a \leqdot b$ to mean $a \lessdot b$ or $a=b$. For any $n\in \mathbb{N}$, denote the poset $[n]$ as the set $\{1, \ldots, n\}$ with the usual order. For any matrix $M$, denote $M(i,j)$ and $M^{\trans}$ to be the $(i,j)$-th entry and transpose of $M$ respectively. If $M$ has real entries, then ${\|M\|}_1$ denotes the sum of the absolute values of all entries in $M$.

Sections \ref{Fountain Code as a Stochastic Process} and \ref{Fountain Matrix} provide a rigorous mathematical framework and the necessary tools, so that in Section \ref{Choosing the Next Probability Distribution}, we can prove that optimal codes have varying probability distributions and give an explicit criterion for optimal code designs.

\section{Fountain Code as a Stochastic Process}\label{Fountain Code as a Stochastic Process}
Let $\mathcal{C}$ be a fountain code with $k$ input symbols $\vec{w}_1, \ldots, \vec{w}_k$ in $\mathbb{F}_2^{\ell}$, and denote $W$ as the $\ell\times k$ matrix $[\vec{w}_1, \ldots, \vec{w}_k]$. For any probability distribution $\mathcal{D}$ on $\mathbb{F}_2^k$, denote $\mathcal{D}(\vec{v})$ as the probability that vector $\vec{v}$ is chosen. Output symbols are generated by the map $\vec{v} \mapsto W\vec{v}$, and we assume the information of $\vec{v}$ is also transmitted, which in practice can be done using a header packet or via some time-synchronization between the source and destination \cite{Raptor}, i.e. we treat the transmission of output symbols as the transmission of column vectors in $\mathbb{F}_2^k$.

For each $n\in \mathbb{N}$, denote $\mathcal{M}_n$ as the set of all $k\times n$ matrices over $\mathbb{F}_2$. Denote $\mathcal{M} = \bigcup_{n\in \mathbb{N}} \mathcal{M}_n$, denote $\mathcal{U}$ as the collection of all subspaces of $\mathbb{F}_2^k$, and denote the poset $\mathcal{U}^{\subseteq}$ as the set $\mathcal{U}$ ordered by set inclusion. For each $r\in \{0,1,\ldots, k\}$, denote $\mathcal{U}_r$ as the collection of all $r$-dimensional subspaces of $\mathbb{F}_2^k$, denote $J_r = |\mathcal{U}_r|$, and denote $K_r = \sum_{i=r}^k J_i$. Also, for any matrix $M \in \mathcal{M}$, denote $u(M)$ as the column space of $M$.

\defn A {\it probability distribution sequence} (abbreviated: {\it p.d.s.}) is a sequence $\{\mathcal{D}_t\}_{t\in \mathbb{N}}$ such that $\mathcal{D}_t$ is a probability distribution on $\mathbb{F}_2^k$ for each $t\in \mathbb{N}$. If $\mathcal{D}_t = \mathcal{D}$ for every $t\in \mathbb{N}$, then we say this p.d.s. is {\it constant}, and by abuse of notation, we breviate this constant sequence simply as $\mathcal{D}$. \defend

\defn A {\it stream of output symbols} with associated p.d.s. $\{\mathcal{D}_t\}_{t\in \mathbb{N}}$ is a sequence of independent discrete random variables $\{X_t\}_{t\in \mathbb{N}}$ such that for each $t\in \mathbb{N}$, we have
\begingroup
\setlength\abovedisplayskip{0.1em}
\setlength\belowdisplayskip{0.3em}
\begin{equation}
X_t: (\mathbb{F}_2^k, 2^{\mathbb{F}_2^k}, \mathcal{D}_t) \to (\mathbb{F}_2^k, 2^{\mathbb{F}_2^k})
\end{equation}
\endgroup
given by the mapping $\vec{v} \mapsto \vec{v}$. A {\it $(k, \{\mathcal{D}_t\}_{t\in \mathbb{N}})$-fountain code} is a set of $k$ input symbols $\{\vec{w}_1, \ldots, \vec{w}_k\}$ together with a stream of output symbols with associated p.d.s. $\{\mathcal{D}_t\}_{t\in \mathbb{N}}$. The special case of a fountain code with constant p.d.s. $\mathcal{D}$ is a $(k, \mathcal{D})$-fountain code, which coincides with the notation in \cite{Raptor}.\defend

For a $(k, \{\mathcal{D}_t\}_{t\in \mathbb{N}})$-fountain code over an erasure channel with erasure probability $\varepsilon$, we can treat erasures as zero vectors transmitted, hence the p.d.s. at the destination is effectively $\{\mathcal{D}_t^*\}_{t\in \mathbb{N}}$, where each $\mathcal{D}_t^*$ is given by $\mathcal{D}_t^*(\vec{v}) = (1-\varepsilon)\mathcal{D}_t(\vec{v})$ for any non-zero $\vec{v} \in \mathbb{F}_2^k$, and $\mathcal{D}_t^*(\vec{0}) = \varepsilon + (1-\varepsilon)\mathcal{D}(\vec{0})$. This means output symbols generated according to $\{\mathcal{D}_t\}_{t\in \mathbb{N}}$, that may be erased with erasure probabilty $\varepsilon$, are equivalent to output symbols generated according to $\{\mathcal{D}_t^*\}_{t\in \mathbb{N}}$ that experience no erasures. Essentially, we have transformed the problem of a channel with erasures to the problem of a channel with no erasures, but with the restriction $\mathcal{D}_t^*(\vec{0}) \geq \varepsilon$ for every $t\in \mathbb{N}$.

\defn Let $\{X_t\}_{t\in \mathbb{N}}$ be a stream of output symbols with associated p.d.s. $\{\mathcal{D}_t\}_{t\in \mathbb{N}}$. For each $t\in \mathbb{N}$, denote $\sigma_t$ as the probability distribution on $\mathcal{M}_t$ given by
\begingroup
\setlength\abovedisplayskip{-0.1em}
\setlength\belowdisplayskip{0.2em}
\begin{equation}\label{sigma-prob}
\sigma_t([\vec{v}_1, \ldots, \vec{v}_t]) = \displaystyle\prod_{i=1}^t \mathcal{D}_i(\vec{v}_i),
\end{equation}
\endgroup
which equals $\Pr(X_1\cdots X_t=[\vec{v}_1, \ldots, \vec{v}_t])$. Define the discrete random variable $G_t = X_1\cdots X_t$ as the concatenation of the first $t$ random variables in the sequence $\{X_t\}_{t\in \mathbb{N}}$, so that
\begingroup
\setlength\abovedisplayskip{0.1em}
\setlength\belowdisplayskip{0.1em}
\begin{equation}
G_t: (\mathcal{M}_t, 2^{\mathcal{M}_t}, \sigma_t) \to (\mathcal{M}_t, 2^{\mathcal{M}_t})
\end{equation}
\endgroup
given by the map $[\vec{v}_1, \ldots, \vec{v}_t] \mapsto [\vec{v}_1, \ldots, \vec{v}_t]$. We say $\{G_t\}_{t\in \mathbb{N}}$ is a {\it generator matrix sequence} with associated p.d.s. $\{\mathcal{D}_t\}_{t\in \mathbb{N}}$. Each random variable $G_t$ is called a {\it generator matrix}.

Also, let $f: \mathcal{M} \to \mathcal{U}$ be the map $M \mapsto u(M)$. For each $t\in \mathbb{N}$, define the discrete random variable
\begingroup
\setlength\abovedisplayskip{0.1em}
\setlength\belowdisplayskip{0.2em}
\begin{equation}
Y_t: (\mathcal{M}_t, 2^{\mathcal{M}_t}, \sigma_t) \to (\mathcal{U}, 2^{\mathcal{U}})
\end{equation}
\endgroup
by the relation $Y_t = f(G_t)$. We call $\{Y_t\}_{t \in \mathbb{N}}$ a {\it generator subspace sequence} with associated p.d.s. $\{\mathcal{D}_t\}_{t\in \mathbb{N}}$.\defend

It is well-known in combinatorics that $\mathcal{U}^{\subseteq}$ is a (finite) complete modular lattice and hence a graded poset, with the dimension of the subspace as the rank function of this poset. In particular, the modularity of $\mathcal{U}^{\subseteq}$ just means
\begingroup
\setlength\abovedisplayskip{0.2em}
\setlength\belowdisplayskip{0.3em}
\begin{equation}\label{modularity}
\dim(U_1) + \dim(U_2) = \dim(U_1 \wedge U_2) + \dim(U_1 \vee U_2)
\end{equation}
\endgroup
for all $U_1, U_2 \in \mathcal{U}^{\subseteq}$. Since $Y_n=u(G_n)=u(X_1\cdots X_n)$ is by definition a random variable representing the span of the columns $X_1, \ldots, X_n$, i.e. the smallest subspace in $\mathcal{U}$ containing subspaces $u(X_1), \ldots, u(X_n)$, we have the identity
\begingroup
\setlength\abovedisplayskip{0.1em}
\setlength\belowdisplayskip{0.3em}
\begin{equation}\label{veee}
Y_n = \bigvee_{i=1}^n u(X_i)
\end{equation}
\endgroup
for all $n\in \mathbb{N}$. The next result then easily follows.

\prop\label{Markov} The generator subspace sequence forms a Markov chain.
\begin{IEEEproof}
Choose any $n\in \mathbb{N}$, and let $U_1, \ldots, U_n$ be arbitrary instances of the random variables $Y_1, \ldots, Y_n$ respectively, such that $\Pr(Y_n = U_n, Y_{n-1} = U_{n-1}, \ldots, Y_1 = U_1) > 0$. Next, choose an arbitrary $U^{\prime} \in \mathcal{U}$. From \eqref{veee}, we have
\begingroup
\setlength\abovedisplayskip{0.1em}
\setlength\belowdisplayskip{0.2em}
\begin{equation}
Y_{n+1} = \left(\bigvee_{i=1}^n u(X_i) \right) \vee u(X_{n+1}) = Y_n \vee u(X_{n+1}),
\end{equation}
\endgroup
so since $Y_1, \ldots, Y_n$ depend on $X_1, \ldots, X_n$, and since $X_{n+1}$ is by definition independent of $X_1, \ldots, X_n$, we get
\begingroup
\setlength\abovedisplayskip{0.1em}
\setlength\belowdisplayskip{0.2em}
\begin{align}
\nonumber & \Pr(Y_{n+1} = U^{\prime}{\mid}Y_n = U_n, \ldots, Y_1 = U_1)\\
\nonumber =& \Pr(Y_n \vee u(X_{n+1}) = U^{\prime}{\mid}Y_n = U_n, \ldots, Y_1 = U_1)\\
\nonumber =& \Pr(Y_n \vee u(X_{n+1}) = U^{\prime}{\mid}Y_n = U_n)\\
=& \Pr(Y_{n+1} = U^{\prime}{\mid}Y_n = U_n).
\end{align}
\endgroup
Consequently, $\{Y_t\}_{t\in \mathbb{N}}$ forms a Markov chain. \end{IEEEproof}

\defn For any $U, U^{\prime} \in \mathcal{U}$ and any probability distribution $\mathcal{D}$ on $\mathbb{F}_2^k$, we define
\begingroup
\setlength\abovedisplayskip{-0.3em}
\setlength\belowdisplayskip{0.1em}
\begin{equation}
\beta_{\mathcal{D}}(U, U^{\prime}) = \begin{cases} \!\!\!\!\!\!\!\displaystyle\sum_{\ \ \ \vec{v} \in U^{\prime} \backslash U} \!\!\!\!\!\!\!\mathcal{D}(\vec{v}), & \text{ if }U \lessdot U^{\prime};\\ \displaystyle\sum_{\vec{v} \in U} \mathcal{D}(\vec{v}), & \text{ if }U^{\prime} = U;\\ 0, & \text{ otherwise.} \end{cases} \tag*{\raisebox{-6ex}{$\IEEEQEDclosed$}}
\end{equation}
\endgroup

\lemma\label{conditional-lemma} Let $\{Y_t\}_{t\in \mathbb{N}}$ be a generator subspace sequence with associated p.d.s. $\{\mathcal{D}_t\}_{t\in \mathbb{N}}$. If $U \in \mathcal{U}$ and $n\in \mathbb{N}$ such that $\Pr(Y_n = U) > 0$, then for any $U^{\prime} \in \mathcal{U}$, we have
\begingroup
\setlength\abovedisplayskip{0.2em}
\setlength\belowdisplayskip{0.5em}
\begin{equation}\label{transMatrix-prob}
\Pr(Y_{n+1}=U^{\prime}{\mid}Y_n=U) = \beta_{\mathcal{D}}(U, U^{\prime}).
\end{equation}
\endgroup

\begin{IEEEproof}
Since $\Pr(Y_n = U) > 0$, there exists some $M\in \mathcal{M}_n$, with $\sigma_n(M) > 0$, such that $u(M) = U$. For any arbitrary $\vec{v} \in \mathbb{F}_2^k$, it follows from \eqref{veee} that
\begingroup
\setlength\abovedisplayskip{0.2em}
\setlength\belowdisplayskip{0.4em}
\begin{equation}
u([M, \vec{v}]) = u(M) \vee u(\vec{v}) = U \vee u(\vec{v}).
\end{equation}
\endgroup
Since $\dim(u(\vec{v})) \leq 1$, we have \eqref{modularity} implies
\begingroup
\setlength\abovedisplayskip{0.3em}
\setlength\belowdisplayskip{0.4em}
\begin{equation}
\dim(u([M, \vec{v}])) = \dim(U \vee u(\vec{v})) \leq \dim(U) + 1,
\end{equation}
\endgroup
hence $U \leqdot u([M, \vec{v}])$, and $\Pr(Y_{n+1} = U^{\prime}{\mid}Y_n = U) \neq 0$ implies $U \leqdot U^{\prime}$. In particular, $U \lessdot u([M, \vec{v}])$ if and only if $\vec{v} \in u([M, \vec{v}]) \backslash U$, and $u([M, \vec{v}]) = U$ if and only if $\vec{v} \in U$. Thus, by considering: 1) $U \lessdot U^{\prime}$; and 2) $U^{\prime} = U$, we sum up the probabilities of the possible vectors and get \eqref{transMatrix-prob}.
\end{IEEEproof}

\section{Fountain Matrix}\label{Fountain Matrix}
Combining Proposition \ref{Markov} and Lemma \ref{conditional-lemma}, we know $\{Y_t\}_{t\in \mathbb{N}}$ is a stationary Markov chain if the corresponding p.d.s. $\{\mathcal{D}_t\}_{t\in \mathbb{N}}$ is constant, which then allows us to use its transition matrix to compute transitional probabilities. In this section, our main goal is to construct the fountain matrix, which is an analogue of the transition matrix in the general non-constant p.d.s. case and an important tool for p.d.s. design analysis.

\defn Let $\mathbb{P}$ be an arbitrary finite poset with $n$ elements. Recall that a refinement of $\mathbb{P}$ is another poset ${\mathbb{P}}^{\prime}$ such that $\mathbb{P} = {\mathbb{P}}^{\prime}$ as sets, and such that for any $x,y\in \mathbb{P}$, we have $x \leq y$ in $\mathbb{P}$ implies $x \leq y$ in ${\mathbb{P}}^{\prime}$. Let ${\mathbb{P}}^*$ be any refinement of $\mathbb{P}$ such that ${\mathbb{P}}^*$ is totally ordered\footnote{Such a refinement is possible, for example by considering the Hasse diagram of $\mathbb{P}$ and refining $\mathbb{P}$ by imposing additional cover relations from left to right. See \cite[Ch. 3]{stanley} for more details.}, and let $\phi: {\mathbb{P}}^* \to [n]$ be the (unique) poset isomorphism corresponding to this refinement. We say $\phi$ is a {\it total-refinement map of $\mathbb{P}$}, and we say ${\mathbb{P}}^*$ is the {\it refinement corresponding to $\phi$}. \defend

\defn Let $\phi$ be a total-refinement map of $\mathcal{U}^{\subseteq}$. Let $n\in \mathbb{N}$, and let $\mathcal{D}, \mathcal{D}_1, \ldots, \mathcal{D}_n$ be probability distributions on $\mathbb{F}_2^k$. We define the {\it $(\mathcal{D}, \phi)$-fountain matrix} as the matrix $T$ given by
\begingroup
\setlength\abovedisplayskip{0.1em}
\setlength\belowdisplayskip{0.2em}
\begin{equation}\label{fountain matrix entry}
T(i,j) = \beta_{\mathcal{D}}(\phi^{{}_{-1}}(i), \phi^{{}_{-1}}(j)).
\end{equation}
\endgroup
Let $T_i$ be the $(\mathcal{D}_i, \phi)$-fountain matrix for each $i\in [n]$, and denote $F_n$ as the matrix product $F_n = T_1\cdots T_n$. We call $F_n$ the {\it $\phi$-fountain product} corresponding to the $n$-tuple $(\mathcal{D}_1, \ldots, \mathcal{D}_n)$. When $n, \phi$ and the corresponding probability distributions are not important, we simply say $T$ is a {\it fountain matrix} and $F_n$ is a {\it fountain product}.\defend

\defn Let $\{\mathcal{D}_t\}_{t\in \mathbb{N}}$ be a p.d.s., and let $\phi$ be any total-refinement map of $\mathcal{U}^{\subseteq}$. For each $n\in \mathbb{N}$, let $T_n$ be the $(\mathcal{D}_n, \phi)$-fountain matrix, and let $F_n$ be the $\phi$-fountain product corresponding to $(\mathcal{D}_1, \ldots, \mathcal{D}_n)$. We then refer to $\{T_t\}_{t\in \mathbb{N}}$ and $\{F_t\}_{t\in \mathbb{N}}$ as the {\it fountain matrix sequence} and {\it fountain product sequence} respectively, each corresponding to $(\{\mathcal{D}_t\}_{t\in \mathbb{N}}, \phi)$.\defend

\defn Let $n\in \mathbb{N}$, let $F$ be a $\phi$-fountain product corresponding to $(\mathcal{D}_1, \ldots, \mathcal{D}_n)$, and let $r, r^{\prime} \in \{0,1,\ldots, k\}$. Let $\mathcal{U}^*$ be the refinement corresponding to $\phi$, let $\mathcal{U}_r^*$ be the induced subposet of $\mathcal{U}^*$ corresponding to set $\mathcal{U}_r$, and let $\phi_r: \mathcal{U}_r^* \to [J_r]$ be the natural (unique) poset isomorphism. Define $\mathcal{U}_{r^{\prime}}^*$ and $\phi_{r^{\prime}}$ analogously. Define the {\it $(r, r^{\prime})$-fountain block of $F$} as the submatrix of $F$ corresponding to the $J_r$ rows in $\{\phi(U): U \in \mathcal{U}_r\}$ and the $J_{r^{\prime}}$ columns in $\{\phi(U) : U \in \mathcal{U}_{r^{\prime}}\}$. We say $\phi_r$ is the {\it $r$-th sub-refinement map} of $F$.\defend

Denoting the $(r, r^{\prime})$-fountain block as $A_{r,r^{\prime}}$, we note that
\begingroup
\setlength\abovedisplayskip{0.2em}
\setlength\belowdisplayskip{0.3em}
\begin{equation}
A_{r,r^{\prime}}(i,j) = F(\phi(\phi_r^{{}_{-1}}(i)), \phi(\phi_{r^{\prime}}^{{}_{-1}}(j))).
\end{equation}
\endgroup
In particular, if $F$ is the $(\mathcal{D}, \phi)$-fountain matrix $T$, then
\begingroup
\setlength\abovedisplayskip{0.2em}
\setlength\belowdisplayskip{0.3em}
\begin{equation}\label{fountain block entry}
A_{r,r^{\prime}}(i,j) = T(\phi(\phi_r^{{}_{-1}}(i)), \phi(\phi_{r^{\prime}}^{{}_{-1}}(j))) = \beta_{\mathcal{D}}(\phi_r^{{}_{-1}}(i), \phi_{r^{\prime}}^{{}_{-1}}(j)).
\end{equation}
\endgroup
Also, since $\dim$ is the rank function of the graded poset $\mathcal{U}^{\subseteq}$, the set $\{\phi(U): U \in \mathcal{U}_r\} \subseteq [K]$ is invariant over all total-refinement maps $\phi$ of $\mathcal{U}^{\subseteq}$, thus we have:
\begingroup
\setlength\abovedisplayskip{0.2em}
\setlength\belowdisplayskip{0.3em}
\begin{equation}\label{indep-choice-of-phi}
{\|A_{r,r^{\prime}}\|}_1 \text{ is independent of the choice of }\phi.
\end{equation}
\endgroup

\thm\label{fountain-matrix-properties} Let $T$ be a $(\mathcal{D}, \phi)$-fountain matrix. Let $r, r^{\prime} \in \{0,1, \ldots, k\}$ be arbitrarily chosen, and denote $A_{r,r^{\prime}}$ as the $(r,r^{\prime})$-fountain block of $T$. We have the following properties: \begin{enumerate}
\item\label{zero matrix item} $A_{r,r^{\prime}}$ is a zero matrix if $r^{\prime} < r$ or $r^{\prime} > r+1$.
\item\label{diagonal block item} $A_{r,r}$ is a diagonal matrix, with
\begingroup
\setlength\abovedisplayskip{0.1em}
\setlength\belowdisplayskip{0.1em}
\begin{equation}
\!\!\!\!\!\!A_{r,r}(m,m) = \!\!\!\!\!\!\sum_{\vec{v} \in \phi_r^{{}_{-1}}(m)} \!\!\!\!\!\mathcal{D}(\vec{v})
\end{equation}
\endgroup
for each $m\in [J_r]$.
\item\label{row sum item} The row sum of each row in $T$ is $1$.
\end{enumerate}

\begin{IEEEproof}
By definition, $\beta_{\mathcal{D}}(U, U^{\prime}) \neq 0$ implies $U \leqdot U^{\prime}$. Since $U \lessdot U^{\prime}$ implies $\rank(U^{\prime}) = \rank(U) + 1$, we get $\beta_{\mathcal{D}}(U, U^{\prime}) \neq 0$ implies $U^{\prime} = U$ or $\rank(U^{\prime}) = \rank(U) + 1$, thus proving property \ref{zero matrix item}, as well as the diagonality of $A_{r,r}$ for every $r\in \{0,1,\ldots, k\}$. Using \eqref{fountain block entry}, we can compute the diagonal entries explicitly, and property \ref{diagonal block item} easily follows.

Choose any $U \in \mathcal{U}$. By definition, row $\phi(U)$ has row sum
\begingroup
\setlength\abovedisplayskip{0.2em}
\setlength\belowdisplayskip{0.2em}
\begin{align}\label{row sum expansion}
\nonumber \sum_{U^{\prime} \in \mathcal{U}} \beta_{\mathcal{D}}(U, U^{\prime}) &= \beta_{\mathcal{D}}(U, U) + \sum_{\substack{U^{\prime} \in \mathcal{U}\\ U \lessdot U^{\prime}}} \beta_{\mathcal{D}}(U, U^{\prime})\\
&= \sum_{\vec{v} \in U} \mathcal{D}(\vec{v}) + \sum_{\substack{U^{\prime} \in \mathcal{U}\\ U \lessdot U^{\prime}}} \sum_{\vec{v} \in U^{\prime} \backslash U} \mathcal{D}(\vec{v}).
\end{align}
\endgroup
We obviously have $\sum_{\vec{v} \in \mathbb{F}_2^k} \mathcal{D}(\vec{v}) = 1$, hence to prove property \ref{row sum item}, it suffices to show that all the summands in the sum in \eqref{row sum expansion} are distinct, and that every vector $\vec{v} \in \mathbb{F}_2^k$ corresponds to exactly one such summand $\mathcal{D}(\vec{v})$ in \eqref{row sum expansion}. For brevity, denote $S$ as the partial sum $\sum_{\vec{v} \in U} \mathcal{D}(\vec{v})$, and for each $U^{\prime} \in \mathcal{U}$ covering $U$, denote $S_{U^{\prime}}$ as the partial sum $\sum_{\vec{v} \in U^{\prime} \backslash U} \mathcal{D}(\vec{v})$. Since $U \cap (U^{\prime}\backslash U) = \emptyset$, the summands in $S$ are distinct from the summands in $S_{U^{\prime}}$ for all $U^{\prime}$ covering $U$.

Now, suppose $U_1^{\prime}, U_2^{\prime} \in \mathcal{U}$ both cover $U$. If there is some $\vec{u} \in \mathbb{F}_2^k \backslash U$ such that $\vec{u} \in U_1^{\prime} \cap U_2^{\prime}$, then $\vec{u} \not\in U$ implies both $U < U \vee u(\vec{u}) \leq U_1^{\prime}$ and $U < U \vee u(\vec{u}) \leq U_2^{\prime}$ in the poset $\mathcal{U}^{\subseteq}$, which forces the equality $U_1^{\prime} = U_2^{\prime}$ as $U_1^{\prime}, U_2^{\prime}$ both cover $U$. This means any pair of distinct partial sums in $\{S_{U^{\prime}} : U^{\prime} \in \mathcal{U}, U \lessdot U^{\prime}\}$ must have distinct summands, thus every summand in \eqref{row sum expansion} appears exactly once. Finally, for any $\vec{v} \in \mathbb{F}_2^k$, we have $\vec{v} \in U \vee u(\vec{v})$. Since $U \leqdot U \vee u(\vec{v})$, it follows that $\mathcal{D}(\vec{v})$ is a summand in \eqref{row sum expansion}, therefore proving property \ref{row sum item}.
\end{IEEEproof}

\lemma\label{marginal} Let $\{Y_t\}_{t\in \mathbb{N}}$ be a generator subspace sequence with associated p.d.s. $\{\mathcal{D}_t\}_{t\in \mathbb{N}}$, let $\phi$ be a total-refinement map of $\mathcal{U}^{\subseteq}$, and let $\{F_t\}_{t\in \mathbb{N}}$ be the corresponding fountain product sequence. Then for any $n\in \mathbb{N}$ and any $U \in \mathbb{U}$, we have
\begingroup
\setlength\abovedisplayskip{0em}
\setlength\belowdisplayskip{0.1em}
\begin{equation}
\Pr(Y_n = U) = F_n(1, \phi(U)).
\end{equation}
\endgroup

\begin{IEEEproof}
This is analogous to the computation of marginal probabilities via the Chapman-Kolmogorov equations (see, e.g., \cite{probability}, \cite{markov-book}), and the consideration of the product of transition matrices of a non-homogenous Markov chain \cite{markov-book}.
\end{IEEEproof}

\cor\label{alpha-lemma} Let $\{G_t\}_{t\in \mathbb{N}}$ be a generator matrix sequence with associated p.d.s. $\{\mathcal{D}_t\}_{t\in \mathbb{N}}$. Let $\phi$ be any total-refinement map of $\mathcal{U}^{\subseteq}$, and let $\{F_t\}_{t\in \mathbb{N}}$ be the corresponding fountain product sequence. For any $n\in \mathbb{N}$ and any arbitrary $r \in \{0,1,\ldots, k\}$, denote $A_{0,r}^{(n)}$ as the $(0,r)$-fountain block of $F_n$. Then for any non-empty subset $S$ of $\{0,1,\ldots, k\}$, we have
\begingroup
\setlength\abovedisplayskip{0.1em}
\setlength\belowdisplayskip{0.2em}
\begin{equation}\label{Pr-rank-of-Gn}
\Pr(\rank(G_n) \in S) = \sum_{r\in S} {\|A_{0,r}^{(n)}\|}_1,
\end{equation}
\endgroup
and this value is independent of the choice of $\phi$.

\begin{IEEEproof}
From \eqref{indep-choice-of-phi}, $\sum_{r\in S} {\|A_{0,r}^{(n)}\|}_1$ is independent of the choice of $\phi$. The rest follows easily from Lemma \ref{marginal}.
\end{IEEEproof}

Using the same notations as above, we define
\begingroup
\setlength\abovedisplayskip{-0.1em}
\setlength\belowdisplayskip{0em}
\begin{equation}
\alpha_{r,n}(\mathcal{D}_1, \ldots, \mathcal{D}_n) = \sum_{j = r}^k {\|A_{0,j}^{(n)}\|}_1
\end{equation}
\endgroup
for $n\in \mathbb{N}$, $r\in \{0,1,\ldots, k\}$. If the context is clear (i.e. when $\{\mathcal{D}_t\}_{t\in \mathbb{N}}$ and $\varepsilon$ are given), then we breviate $\alpha_{r,n}(\mathcal{D}_1^*, \ldots, \mathcal{D}_n^*)$ as $\alpha_{r,n}$. Corollary \ref{alpha-lemma} then tells us $\alpha_{r,n}$ is the probability of the generator matrix $G_n$ having rank $\geq r$ at the destination.

\section{Choosing the Next Probability Distribution}\label{Choosing the Next Probability Distribution}
Let $N\in \mathbb{N}$, and suppose we are given $N$ probability distributions $\mathcal{D}_1, \ldots, \mathcal{D}_N$ on $\mathbb{F}_2^k$, which form the first $N$ probability distributions of some p.d.s. design at the source. To minimize overhead, we then need to decide on the next distribution $\mathcal{D}_{N+1}$ in the p.d.s. so that $\alpha_{r,N+1}$ is maximized.

\defn Let $n\in \mathbb{N}$, and let $\mathcal{D}_1, \ldots, \mathcal{D}_n$ be any $n$ probability distributions on $\mathbb{F}_2^k$. Let $\phi$ be any total-refinement map of $\mathcal{U}^{\subseteq}$, and let $F$ be the $\phi$-fountain product corresponding to the $n$-tuple $(\mathcal{D}_1, \ldots, \mathcal{D}_n)$. For each $t,i,j \in \{0,1,\ldots, k\}$, denote $\phi_t$ as the $t$-th sub-refinement map of $F$, and denote $A_{i,j}$ as the $(i,j)$-fountain block of $F$. For any $r\in \{0,1,\ldots, k\}$ and any non-zero $\vec{v}\in \mathbb{F}_2^k$, we define 
\begingroup
\setlength\abovedisplayskip{-0.1em}
\setlength\belowdisplayskip{-0.1em}
\begin{equation}
\gamma_{r,n}(\vec{v}|{}\mathcal{D}_1, \ldots, \mathcal{D}_n) = \sum_{\substack{U: U\in \mathcal{U}_r\\ \vec{v}\in U}} \!\!\![A_{0,r}(1, \phi_r(U))].
\end{equation}
\endgroup
By default, set the empty sum $\gamma_{0,n}(\vec{v}|{}\mathcal{D}_1, \ldots, \mathcal{D}_n)$ to be $0$.

Define $\gamma_{r,n}^{\min}(\mathcal{D}_1,\ldots, \mathcal{D}_n)$ [abbreviated: $\gamma_{r,n}^{\min}$ if context is clear] as the minimum value of $\gamma_{r,n}(\vec{v}|{}\mathcal{D}_1, \ldots, \mathcal{D}_n)$, where $\vec{v}$ varies over all non-zero vectors in $\mathbb{F}_2^k$. Also, define
\begingroup
\setlength\abovedisplayskip{0.1em}
\setlength\belowdisplayskip{-0.4em}
\begin{equation}
\Gamma_{r,n}^{\min}(\mathcal{D}_1,\ldots, \mathcal{D}_n) = \displaystyle\argmin_{\text{\smash{$\vec{v}\!\in \!\mathbb{F}_2^k\backslash\{\vec{0}\}$}}} \gamma_{r,n}(\vec{v}|\mathcal{D}_1, \ldots, \mathcal{D}_n),
\end{equation}
\endgroup
{{ }\defend}
\thm\label{next-distributionThm} Let $n\in \mathbb{N}$ and let $\mathcal{D}_1, \ldots, \mathcal{D}_n$ be probability distributions on $\mathbb{F}_2^k$. Let $\phi$ be a total-refinement map of $\mathcal{U}^{\subseteq}$, let $F$ be the $\phi$-fountain product corresponding to $(\mathcal{D}_1, \ldots, \mathcal{D}_n)$, and for each $i,j\in \{0,1,\ldots, k\}$, denote $A_{i,j}$ as the $(i,j)$-fountain block of $F$. For any $t\in \{0,1,\ldots, k\}$, $\vec{v} \in \mathbb{F}_2^k\backslash\{\vec{0}\}$, we breviate $\gamma_{t,n}(\vec{v}|{}\mathcal{D}_1, \ldots, \mathcal{D}_n)$ as $\gamma_{t,n}(\vec{v})$. Then, for any $r\in [k]$ and any probability distribution $\mathcal{D}$ on $\mathbb{F}_2^k$, we have
\begingroup
\setlength\abovedisplayskip{0.1em}
\setlength\belowdisplayskip{0.3em}
\begin{equation}\label{upper-bound-statement}
\alpha_{r,n+1}(\mathcal{D}_1, \ldots, \mathcal{D}_n, \mathcal{D}) = \alpha_{r,n}(\mathcal{D}_1, \ldots, \mathcal{D}_n) + C,
\end{equation}
\endgroup
where $C$ is given by
\begingroup
\setlength\abovedisplayskip{0em}
\setlength\belowdisplayskip{0em}
\begin{equation}
C = (1 - \mathcal{D}(\vec{0})){\|A_{0,r-1}\|}_1 - \Bigl(\sum_{\vec{v}\in\mathbb{F}_2^k\text{\rlap{$\backslash\{\vec{0}\}$}}} \gamma_{r-1,n}(\vec{v}) \mathcal{D}(\vec{v})\Bigr). \label{C-exp}
\end{equation}
\endgroup

\begin{IEEEproof}
See appendix.
\end{IEEEproof}

\cor\label{distribution-bound}
Let $n\in \mathbb{N}$ and let $\mathcal{D}_1, \ldots, \mathcal{D}_n$ be probability distributions on $\mathbb{F}_2^k$ with erasure probability $\varepsilon$. Let $\phi$ be a total-refinement map of $\mathcal{U}^{\subseteq}$, and denote $A_{0,j}$ as the $(0,j)$-fountain block of the $\phi$-fountain product corresponding to $(\mathcal{D}_1^*, \ldots, \mathcal{D}_n^*)$. Then, for any $r\in [k]$ and any probability distribution $\mathcal{D}$ on $\mathbb{F}_2^k$ satisfying $\mathcal{D}^*(\vec{0}) = \varepsilon^{\prime} \geq \varepsilon$, we have
\begingroup
\setlength\abovedisplayskip{0.2em}
\setlength\belowdisplayskip{0.2em}
\begin{equation}
\alpha_{r,n+1}(\mathcal{D}_1^*, \ldots, \mathcal{D}_n^*, \mathcal{D}^*) \leq \alpha_{r,n}(\mathcal{D}_1^*, \ldots, \mathcal{D}_n^*) + C^{\prime},\label{upper-bound-next-dist}
\end{equation}
\endgroup
where $C^{\prime}$ is a non-negative value given by
\begingroup
\setlength\abovedisplayskip{0.1em}
\setlength\belowdisplayskip{0.1em}
\begin{equation}
C^{\prime} = (1 - \varepsilon^{\prime})({\|A_{0,r-1}\|}_1 - \gamma_{r-1,n}^{\min}(\mathcal{D}_1^*,\ldots,\mathcal{D}_n^*)). \label{C-2-exp}
\end{equation}
\endgroup
Breviating $\Gamma_{r-1,n}^{\min}(\mathcal{D}_1^*, \ldots, \mathcal{D}_n^*)$ as $\Gamma_{r-1,n}^{\min}$, equality holds in \eqref{upper-bound-next-dist} if and only if $\mathcal{D}$ satisfies
\begingroup
\setlength\abovedisplayskip{0em}
\setlength\belowdisplayskip{0.1em}
\begin{equation}\label{Gamma-lower-bound}
\sum_{\vec{v} \in \Gamma_{r-1,n}^{\min}} \!\!\!\!\!\mathcal{D}^*(\vec{v}) = 1-\varepsilon^{\prime}.
\end{equation}
\endgroup

\begin{IEEEproof}
By Theorem \ref{next-distributionThm}, it suffices to prove that
\begingroup
\setlength\abovedisplayskip{0em}
\setlength\belowdisplayskip{0.1em}
\begin{equation}\label{gamma-ineq-want-to-prove}
(1 - \varepsilon^{\prime})\gamma_{r-1,n}^{\min} \leq \!\!\!\!\!\!\sum_{\vec{v}\in \mathbb{F}_2^k\backslash\{\vec{0}\}} \!\!\!\!\!\![\gamma_{r-1,n}(\vec{v}) \mathcal{D}^*(\vec{v})]
\end{equation}
\endgroup
Note that by the definition of $\gamma_{r-1,n}(\vec{v})$, we have
\begingroup
\setlength\abovedisplayskip{0.1em}
\setlength\belowdisplayskip{0.2em}
\begin{equation}\label{gamma-inequality-nonneg}
0 \leq \gamma_{r-1,n}^{\min} \leq \gamma_{r-1,n}(\vec{v}) \leq {\|A_{0,r-1}\|}_1
\end{equation}
\endgroup
for all $\vec{v}\in \mathbb{F}_2^k\backslash\{\vec{0}\}$. Treating $\{\mathcal{D}^*(\vec{v}) : \vec{v}\in \mathbb{F}_2^k, \vec{v} \neq \vec{0}\}$ as $2^k-1$ non-negative real-valued variables subject to the constraint
\begingroup
\setlength\abovedisplayskip{0em}
\setlength\belowdisplayskip{0.2em}
\begin{equation}\label{constraint-on-nonzero-vec}
\sum_{\vec{v}\in \mathbb{F}_2^k\backslash\{\vec{0}\}} \!\!\!\!\!\mathcal{D}^*(\vec{v}) = 1 - \varepsilon^{\prime},
\end{equation}
\endgroup
we see that \eqref{gamma-inequality-nonneg} and \eqref{constraint-on-nonzero-vec} imply \eqref{gamma-ineq-want-to-prove}. The lower bound in \eqref{gamma-ineq-want-to-prove} is attained only if $\mathcal{D}^*(\vec{v}) \neq 0$ implies $\vec{v}\in\Gamma_{r-1,n}^{\min}$, hence \eqref{constraint-on-nonzero-vec} yields \eqref{Gamma-lower-bound}. Finally, $C^{\prime}\geq 0$ follows trivially from \eqref{gamma-inequality-nonneg}.
\end{IEEEproof}

\cor\label{complete-answer-next-best-dist} Let $n\in \mathbb{N}$, and let $n$ probability distributions $\mathcal{D}_1, \ldots, \mathcal{D}_n$ on $\mathbb{F}_2^k$ corresponding to erasure probability $\varepsilon$ be given. Then for any $r\in [k]$, and any probability distribution $\mathcal{D}$ with erasure probability $\varepsilon$, we have $\alpha_{r,n+1}(\mathcal{D}_1^*, \ldots, \mathcal{D}_n^*, \mathcal{D}^*)$ is maximized if and only if $\mathcal{D}$ satisfies the condition:
\begingroup
\setlength\abovedisplayskip{0em}
\setlength\belowdisplayskip{0.2em}
\begin{equation}
\mathcal{D}(\vec{v}) \neq 0 \Rightarrow \vec{v} \in \Gamma_{r-1,n}^{\min}(\mathcal{D}_1^*, \ldots, \mathcal{D}_n^*).
\end{equation}
\endgroup

\begin{IEEEproof}
By Corollary \ref{distribution-bound}, since $\mathcal{D}_1, \ldots, \mathcal{D}_n$ are given implies $({\|A_{0,r-1}\|}_1 - \gamma_{r-1,n}^{\min}(\mathcal{D}_1^*,\ldots,\mathcal{D}_n^*))$ in \eqref{C-2-exp} is fixed, the upper bound in \eqref{upper-bound-next-dist} is independent of the choice of $\mathcal{D}^*(\vec{v})$ for non-zero $\vec{v}$. So to maximize $C^{\prime}$ in \eqref{C-2-exp}, we need to minimize $\mathcal{D}^*(\vec{0})$, i.e. set $\mathcal{D}^*(\vec{0}) = \varepsilon$, and the assertion easily follows.
\end{IEEEproof}
\begin{figure}%
\centering%
\includegraphics[width=3.45in]{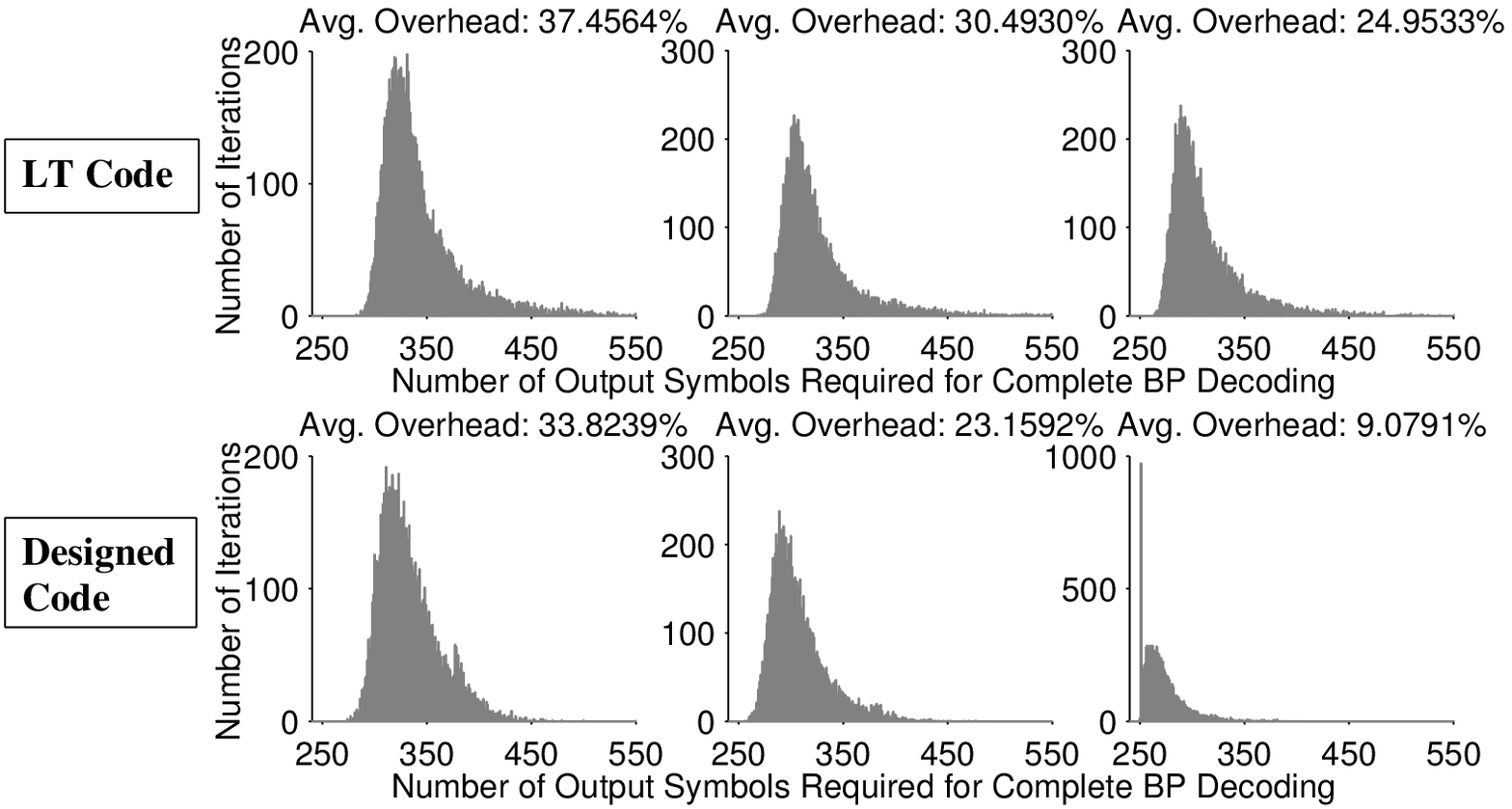}%
\caption{Comparison between LT code and designed code, each with $k=250$ input symbols. Erasure probabilities from left to right: $10\%$, $5\%$, $1\%$.}%
\label{fig:comparisons}%
\end{figure}%
Corollary \ref{complete-answer-next-best-dist} serves as a criterion for choosing the next probability distribution $\mathcal{D}$ in a p.d.s. with some initial probability distributions already determined. In particular, the design of the next distribution $\mathcal{D}$ is independent of $\varepsilon$. Starting with any $\mathcal{D}_1$ satisfying $\mathcal{D}_1(\vec{0}) = 0$, the recursive use of Corollary \ref{complete-answer-next-best-dist} then serves as a criterion for optimal code designs.

Note also that for p.d.s. $\{\mathcal{D}_t\}_{t\in \mathbb{N}}$, if $\mathcal{D}_{n+1}$ is fixed, then by Corollary \ref{distribution-bound}, $\alpha_{r,n+1}$ is maximized if $\gamma_{r-1,n}^{\min}$ is minimized. So for $\{\mathcal{D}_t\}_{t\in \mathbb{N}}$ to be optimal, we must have $\gamma_{r-1,1}^{\min} = 0$, forcing $\gamma_{r-1,2}^{\min}= 0$, hence we get $\mathcal{D}_1 \neq \mathcal{D}_2$. Consequently, optimal codes necessarily have non-constant p.d.s. designs. In particular, such designs are `universal', i.e. they are optimal for erasure channels with arbitrary erasure probabilities, so a constant p.d.s. is not required for the `universality' property.

Following notations in \cite{LTcodes}, the robust soliton distribution depends on parameters $k, c, \delta$, which we denote as $\mu_{k,c,\delta}$. Now, consider the $(k, \{\mathcal{D}_t\}_{t\in \mathbb{N}})$-fountain code $\mathcal{C}^{\prime}$ such that
\begingroup
\setlength\abovedisplayskip{0.2em}
\setlength\belowdisplayskip{0.2em}
\begin{equation}
G_k = \left[\text{
\begin{tabular}{cccccc}
\cline{1-3} \cline{5-5}
\multicolumn{1}{|c|}{\multirow{5}{*}{$\vec{u}${}\rlap{${}_{1}$}}} &
\multicolumn{1}{|c|}{\multirow{4}{*}{$\vec{u}${}\rlap{${}_{2}$}}} &
\multicolumn{1}{|l|}{\multirow{3}{*}{$\vec{u}${}\rlap{${}_{3}$}}} &
$\cdots$ & \multicolumn{1}{|c|}{${}_{\vec{u}_{k-}}${}\rlap{${}_{{}_{1}}$}} & $1$ \\ \cline{5-5}
\multicolumn{1}{|c|}{} & \multicolumn{1}{|c|}{} & \multicolumn{1}{|c|}{} & {} & $1$ & $0$\\
\multicolumn{1}{|c|}{} & \multicolumn{1}{|c|}{} & \multicolumn{1}{|c|}{} & $\iddots$ &    $0$   &    $0$  \\ \cline{3-3}
\multicolumn{1}{|c|}{} & \multicolumn{1}{|c|}{} & $1$ & $\iddots$ & $\vdots$ & $\vdots$\\ \cline{2-2}
\multicolumn{1}{|c|}{} &$1$ &$0$ & $\cdots$  &    $0$   &    $0$  \\ \cline{1-1}
$1$&$0$ &$0$ & $\cdots$ &     $0$    &     $0$
\end{tabular}
} \right],
\end{equation}
\endgroup
where for each $t\in[k-1]$ and some initial $d_t\in[k-t]$ chosen according to degree distribution $\mu_{k-t,c,\delta}$, we have $\vec{u}_t \in \mathbb{F}_2^{k-t}$ is a vector of Hamming weight $d_t-1$ that is chosen uniformly at random. This description uniquely determines $\mathcal{D}_1, \ldots, \mathcal{D}_k$, which are clearly distinct. Furthermore, we fix $\mathcal{D}_{k+1}, \ldots, \mathcal{D}_{k+\lfloor\frac{k}{2}\rfloor}$ to be probability distributions induced by the degree distribution $\mu_{k,c,\delta}$, and we define subsequent probability distributions recursively: $\mathcal{D}_{n+\lfloor \frac{3k}{2}\rfloor} = \mathcal{D}_n$ for all $n\in \mathbb{N}$. Using Corollary \ref{complete-answer-next-best-dist}, we can check that $\alpha_{r,n}(\mathcal{D}_1^*, \ldots, \mathcal{D}_n^*)$ is maximized for all $r,n\in [k]$ and any erasure probability $\varepsilon$, hence the first $k$ probability distributions of $\mathcal{C}^{\prime}$ are optimal.

Using belief propagation (BP) decoding and \mbox{parameters $k=$} $250$, $c=0.03$, $\delta=0.5$, we run simulations for the \mbox{LT code} and $\mathcal{C}^{\prime}$. Each histogram in Fig. \ref{fig:comparisons} represents $10000$ iterations, and we see that $\mathcal{C}^{\prime}$ has a lower overhead than the LT code, e.g. $9.70\%$, $24.05\%$, $63.62\%$ lower for $\varepsilon = 10\%$, $5\%$, $1\%$ respectively. Note that other code designs with further overhead reduction are possible and will be elaborated in future work.

\section{Conclusion}\label{Conclusion}
In this paper, we applied probability theory and the theory of posets, and showed that optimal fountain codes must have non-constant p.d.s. designs. A criterion for optimal code designs has also been derived. Simulations show an overhead reduction when the probability distributions are varied, hence our theory of optimal p.d.s. designs has immense significance for fountain codes, promising codes with low overhead.

\appendix[Proof of Theorem \ref{next-distributionThm}]\label{proof-of-next-distributionThm}
Denote $\varepsilon = \mathcal{D}(\vec{0})$. For each $t,i,j\in \{0,1,\ldots, k\}$, denote $\phi_t$ as the $t$-th sub-refinement map of $F$, and denote $A_{i,\geqslant r}$ and $A_{\geqslant r,j}$ as the concatenations of matrices:
\begingroup
\setlength\abovedisplayskip{0.2em}
\setlength\belowdisplayskip{0.3em}
\begin{align}
A_{i,\geqslant r} &= [A_{i,r}, A_{i,r+1}, \ldots, A_{i,k}];\\
A_{\geqslant r, j} &= [A_{r,j}^{\trans}, A_{r+1,j}^{\trans}, \ldots, A_{k,j}^{\trans}]^{\trans}.
\end{align}
\endgroup
Also, we denote $A_{\geqslant r, \geqslant r}$ by the two equivalent expressions
\begingroup
\setlength\abovedisplayskip{0.2em}
\setlength\belowdisplayskip{0.3em}
\begin{align}
A_{\geqslant r, \geqslant r} &= [A_{\geqslant r, r}, A_{\geqslant r, r+1}, \ldots, A_{\geqslant r,k}]\\
&= [A_{r,\geqslant r}^{\trans}, A_{r+1,\geqslant r}^{\trans}, \ldots, A_{k,\geqslant r}^{\trans}]^{\trans}.
\end{align}
\endgroup
We can then partition $F$ into $(r+1)^2$ blocks so that
\begingroup
\setlength\abovedisplayskip{0.1em}
\setlength\belowdisplayskip{0.2em}
\begin{equation}\label{block-structure}
F = \begin{bmatrix}
A_{0,0}&A_{0,1}&\cdots&A_{0,r-1}&A_{0,\geqslant{r}}\\
A_{1,0}&A_{1,1}&\cdots&A_{1,r-1}&A_{1,\geqslant{r}}\\
\vdots&\vdots&\ddots&\vdots&\vdots\\
A_{r-1,0}&A_{r-1,1}&\cdots&A_{r-1,r-1}&A_{r-1,\geqslant{r}}\\
A_{\geqslant{r},0}&A_{\geqslant{r},1}&\cdots&A_{\geqslant{r},r-1}&A_{\geqslant{r},\geqslant{r}}\\
\end{bmatrix}.
\end{equation}
\endgroup
\quad Next, denote $T$ as the $(\mathcal{D}, \phi)$-fountain matrix, and denote $A_{i,j}^{\prime}$ as the $(i,j)$-fountain block of $T$. Defining $A_{i,\geqslant{r}}^{\prime}$, $A_{\geqslant{r},j}^{\prime}$ and $A_{\geqslant{r},\geqslant{r}}^{\prime}$ analogously, we can partition $T$ into $(r+1)^2$ blocks analogous to \eqref{block-structure}. The product $FT$ is the $\phi$-fountain product corresponding to $(\mathcal{D}_1,\ldots,\mathcal{D}_n,\mathcal{D})$, thus we get
\begingroup
\setlength\abovedisplayskip{-0.2em}
\setlength\belowdisplayskip{0.4em}
\begin{equation}\label{first-eqn-for-alpha-r-n+1}
\alpha_{r,n+1}(\mathcal{D}_1,\ldots,\mathcal{D}_n,\mathcal{D}) = {\biggl\|\biggl(\sum_{i=0}^{r-1}A_{0,i}A_{i,\geqslant{r}}^{\prime}\biggl) + A_{0,\geqslant{r}}A_{\geqslant{r},\geqslant{r}}^{\prime}\biggr\|}_1.
\end{equation}
\endgroup
Using Theorem \ref{fountain-matrix-properties}\ref{zero matrix item}, \eqref{first-eqn-for-alpha-r-n+1} reduces to
\begingroup
\setlength\abovedisplayskip{-0.1em}
\setlength\belowdisplayskip{0em}
\begin{equation}\label{second-eqn-for-alpha-r-n+1}
\alpha_{r,n+1}(\mathcal{D}_1,\ldots,\mathcal{D}_n,\mathcal{D}) = {\|A_{0,r-1}A_{r-1,\geqslant{r}}^{\prime}\|}_1 + {\|A_{0,\geqslant{r}}A_{\geqslant{r},\geqslant{r}}^{\prime}\|}_1.
\end{equation}
\endgroup
\quad Denote $B_r$ and $B_r^{\prime}$ for $A_{0,r-1}A_{r-1,\geqslant{r}}^{\prime}$ and $A_{0,\geqslant{r}}A_{\geqslant{r},\geqslant{r}}^{\prime}$ respectively. Then for each $j\in [K_r]$, we have
\begingroup
\setlength\abovedisplayskip{0em}
\setlength\belowdisplayskip{0.1em}
\begin{equation}
B_r(1,j) = \sum_{i=1}^{J_{r-1}} [A_{0,r-1}(1,i)][A_{r-1,\geqslant{r}}^{\prime}(i,j)],
\end{equation}
\endgroup
hence taking the norm, we get
\begingroup
\setlength\abovedisplayskip{-0.1em}
\setlength\belowdisplayskip{0.1em}
\begin{align}
\nonumber \!{\|B_r\|}_1 &= \sum_{j=1}^{K_r} B_r(1,j) = \sum_{j=1}^{K_r} \sum_{i=1}^{J_{r-1}}[A_{0,r-1}(1,i)]\text{\rlap{$[A_{r-1,\geqslant{r}}^{\prime}(i,j)]$}}\\
&= \sum_{i=1}^{J_{r-1}} \biggl[[A_{0,r-1}(1,i)] \biggl(\sum_{j=1}^{K_r} [A_{r-1,\geqslant{r}}^{\prime}(i,j)]\biggr) \biggr].\label{br-eqn}
\end{align}
\endgroup
Similarly, we also get
\begingroup
\setlength\abovedisplayskip{0em}
\setlength\belowdisplayskip{0.1em}
\begin{equation}\label{brprime-eqn}
{\|B_r^{\prime}\|}_1 = \sum_{i=1}^{K_r} \biggl[[A_{0,\geqslant{r}}(1,i)] \biggl(\sum_{j=1}^{K_r} [A_{\geqslant{r},\geqslant{r}}^{\prime}(i,j)]\biggr) \biggr].
\end{equation}
\endgroup
Theorem \ref{fountain-matrix-properties}\ref{zero matrix item} tells us that $A_{\geqslant{r},t}^{\prime}$ is a zero matrix for each $t\in \{0,\ldots,r-1\}$, so Theorem \ref{fountain-matrix-properties}\ref{row sum item} implies $\sum_{j=1}^{K_r} [A_{\geqslant{r},\geqslant{r}}^{\prime}(i,j)] = 1$ for each $i\in [K_r]$, thus \eqref{brprime-eqn} reduces to
\begingroup
\setlength\abovedisplayskip{0.4em}
\setlength\belowdisplayskip{0.8em}
\begin{equation}\label{brprime-identity}
\!\!{\|B_r^{\prime}\|}_1 \!= \!\text{\smash{$\displaystyle\sum_{i=1}^{K_r}$}}[A_{0,\geqslant{r}}(1,i)] \!= \!{\|A_{0,\geqslant{r}}\|}_1 \!= \!\alpha_{r,n}(\mathcal{D}_1, \ldots, \text{\rlap{$\mathcal{D}_n).$}}
\end{equation}
\endgroup
\quad Similarly, Theorem \ref{fountain-matrix-properties}\ref{zero matrix item} yields $A_{r-1,j}^{\prime}$ is a zero matrix for each $j\in \{0,\ldots,r-2\}$ (if any). Since Theorem \ref{fountain-matrix-properties}\ref{diagonal block item} tells us $A_{r-1,r-1}^{\prime}$ is a diagonal matrix, with
\begingroup
\setlength\abovedisplayskip{-0.2em}
\setlength\belowdisplayskip{0.1em}
\begin{equation}
A_{r-1,r-1}^{\prime}(i,i) = \!\!\!\!\sum_{\vec{v} \in \phi_{r-1}^{-1}\text{\rlap{$(i)$}}}\!\!\!\mathcal{D}(\vec{v})
\end{equation}
\endgroup
for each $i \in [J_{r-1}]$, it then follows from Theorem \ref{fountain-matrix-properties}\ref{row sum item} that
\begingroup
\setlength\abovedisplayskip{0em}
\setlength\belowdisplayskip{-0.2em}
\begin{equation}
\sum_{j=1}^{K_r} [A_{r-1,\geqslant{r}}^{\prime}(i,j)] = 1 - \biggl(\sum_{\vec{v} \in \phi\text{\rlap{${}_{r-1}^{-1}(i)$}}} \mathcal{D}(\vec{v})\biggr)
\end{equation}
\endgroup
for each $i \in [J_{r-1}]$. Also, for any $t\in \{0,1,\ldots, k\}$, we have
\begingroup
\setlength\abovedisplayskip{0.1em}
\setlength\belowdisplayskip{0.1em}
\begin{align}
\nonumber & \ \ \sum_{U\in \mathcal{U}_t} \Bigl([A_{0,t}(1,\phi_{t}(U))]\sum_{\smash{\substack{\raisebox{0em}[0.3em][0.3em]{\phantom{a}}\\ \vec{v}\in U\\ \vec{v} \neq \vec{0}}}} \mathcal{D}(\vec{v})\Bigr)\\
\nonumber =& \sum_{\vec{v} \in \mathbb{F}_2^k\backslash\{\vec{0}\}} \Bigl[\Bigl(\!\!\!\sum_{\substack{U: U\in \mathcal{U}_t\\{\text{\raisebox{-0.2em}[0pt][0pt]{$\vec{v}\in U$}}}}} \!\![A_{0,t}(1, \phi_t(U))] \Bigr) \mathcal{D}(\vec{v})\Bigr]\\
=& \sum_{\vec{v} \in \mathbb{F}_2^k\backslash\{\vec{0}\}} \!\!\!\!\gamma_{t,n}(\vec{v})\mathcal{D}(\vec{v}).\label{count-in-diff-way-identity}
\end{align}
\endgroup
Since $\mathcal{U}_{r-1} = \{\phi_{(r-1)}^{{}_{-1}}(i): i\in [J_{r-1}]\}$, we have \eqref{br-eqn} yields
\begingroup
\setlength\abovedisplayskip{-0.1em}
\setlength\belowdisplayskip{0.1em}
\begin{align}
\nonumber {\|B_r\|}_1 =& \!\!\sum_{\!\!U\in \mathcal{U}_{r-\text{\rlap{$1$}}}} \Bigl[[A_{0,r-1}(1,\phi_{r-1}(U))]\Bigl(1 - \sum_{\vec{v}\in U} \mathcal{D}(\vec{v})\Bigr)\Bigr]\\
\nonumber =& \sum_{U\in \mathcal{U}_{r\text{\rlap{$-1$}}}} \!\![(1 - \varepsilon) [A_{0,r-1}(1,\phi_{r-1}(U))]]\\
\nonumber &- \!\!\!\sum_{U\in \mathcal{U}_{r-1}} \!\!\!\!\Bigl[[A_{0,r-1}(1,\phi_{r-1}(U))]\sum_{\smash{\substack{\raisebox{0em}[0.3em][0.3em]{\phantom{a}}\\ \vec{v}\in U\\ \vec{v} \neq \vec{0}}}} \mathcal{D}(\vec{v})\Bigr]\\
=&\ (1 - \varepsilon){\|A_{0,r-1}\|}_1 - \Bigl(\sum_{\vec{v}\in\mathbb{F}_2^k\text{\rlap{$\backslash\{\vec{0}\}$}}} \gamma_{r-1,n}(\vec{v}) \mathcal{D}(\vec{v})\Bigr), \label{Br-second-part}
\end{align}
\endgroup
where the last equality follows from \eqref{count-in-diff-way-identity}. Finally, combining \eqref{second-eqn-for-alpha-r-n+1}, \eqref{brprime-identity} and \eqref{Br-second-part}, we get \eqref{upper-bound-statement} as claimed.

\bibliographystyle{IEEEtran}
\bibliography{IEEEabrv,coding_biblio}

\begin{thebibliography}{10}
\providecommand{\url}[1]{#1}
\csname url@samestyle\endcsname
\providecommand{\newblock}{\relax}
\providecommand{\bibinfo}[2]{#2}
\providecommand{\BIBentrySTDinterwordspacing}{\spaceskip=0pt\relax}
\providecommand{\BIBentryALTinterwordstretchfactor}{4}
\providecommand{\BIBentryALTinterwordspacing}{\spaceskip=\fontdimen2\font plus
\BIBentryALTinterwordstretchfactor\fontdimen3\font minus
  \fontdimen4\font\relax}
\providecommand{\BIBforeignlanguage}[2]{{%
\expandafter\ifx\csname l@#1\endcsname\relax
\typeout{** WARNING: IEEEtran.bst: No hyphenation pattern has been}%
\typeout{** loaded for the language `#1'. Using the pattern for}%
\typeout{** the default language instead.}%
\else
\language=\csname l@#1\endcsname
\fi
#2}}
\providecommand{\BIBdecl}{\relax}
\BIBdecl

\bibitem{LTcodes}
M.~Luby, ``{LT} codes,'' in \emph{Proc. 43rd Annu. {IEEE} Symp. Foundations of
  Computer Science {(FOCS)} '02}, Vancouver, {BC}, Canada, Nov. 2002, pp.
  271--280.

\bibitem{fountain}
J.~W. Byers, M.~Luby, M.~Mitzenmacher, and A.~Rege, ``A digital fountain
  approach to reliable distribution of bulk data,'' in \emph{Proc. {ACM}
  {SIGCOMM} '98}, Vancouver, {BC}, Canada, Sep. 1998, pp. 56--67.

\bibitem{Raptor}
A.~Shokrollahi, ``Raptor codes,'' \emph{{IEEE} Trans. Inf. Theory}, vol.~52,
  no.~6, pp. 2551--2567, Jun. 2006.

\bibitem{RT-jnl}
A.~Beimel, S.~Dolev, and N.~Singer, ``{RT} oblivious erasure correcting,''
  \emph{{IEEE/ACM} Trans. Netw.}, vol.~15, no.~6, pp. 1321--1332, Dec. 2007.

\bibitem{shifting2}
A.~Hagedorn, S.~Agarwal, D.~Starobinski, and A.~Trachtenberg, ``Rateless coding
  with feedback,'' in \emph{Proc. 28th {IEEE} Conf. Comput. Commun.
  {(INFOCOMM)} '09}, Rio de Janeiro, Brazil, Apr. 2009, pp. 1791--1799.

\bibitem{systematic}
T.~Nguyen, L.~Yang, and L.~Hanzo, ``Systematic luby transform codes and their
  soft decoding,'' in \emph{Proc. {IEEE} Workshop on Signal Process. Syst.
  '07}, Shanghai, China, Oct. 2007, pp. 67--72.

\bibitem{reconfigurable-jnl}
N.~Bonello, R.~Zhang, S.~Chen, and L.~Hanzo, ``Reconfigurable rateless codes,''
  \emph{{IEEE} Trans. Wireless Commun.}, vol.~8, no.~11, pp. 5592--5600, Nov.
  2009.

\bibitem{probability}
P.~Billingsley, \emph{Probability and Measure}.\hskip 1em plus 0.5em minus
  0.4em\relax Wiley-Interscience, 1995.

\bibitem{stanley}
R.~P. Stanley, \emph{Enumerative combinatorics}, ser. Cambridge Studies in
  Advanced Mathematics.\hskip 1em plus 0.5em minus 0.4em\relax Cambridge
  University Press, 1999, vol.~1.

\bibitem{markov-book}
E.~Seneta, \emph{Non-negative Matrices and Markov Chains}, ser. Springer Series
  in Statistics.\hskip 1em plus 0.5em minus 0.4em\relax Springer, 2006.

\end{thebibliography}

\end{document}